\documentclass[12pt]{iopart}
\usepackage{amssymb}
\begin{document}

\renewcommand{\theequation}{\arabic{section}.\arabic{equation}}

\font\twozero=cmr10 at 20pt \font\oneeight=cmr10 at 18pt
\font\onefour=cmr10 at 14pt \font\larl=cmr10 at 24pt
\newcommand{\vT}{\vphantom{\mbox{\twozero I}}}
\newcommand{\vTm}{\vphantom{\mbox{\oneeight I}}}
\newcommand{\vTs}{\vphantom{\mbox{\onefour I}}}
\newcommand{\vTb}{\vphantom{\mbox{\larl I}}}

\def\a{\alpha}
\def\b{\beta}
\def\d{\delta}
\def\e{\epsilon}
\def\g{\gamma}
\def\h{\mathfrak{h}}
\def\k{\kappa}
\def\l{\lambda}
\def\o{\omega}
\def\p{\wp}
\def\r{\rho}
\def\t{\tau}
\def\s{\sigma}
\def\z{\zeta}
\def\x{\xi}
\def\V={{{\bf\rm{V}}}}
 \def\A{{\cal{A}}}
 \def\B{{\cal{B}}}
 \def\C{{\cal{C}}}
 \def\D{{\cal{D}}}
\def\G{\Gamma}
\def\K{{\cal{K}}}
\def\O{\Omega}
\def\R{\bar{R}}
\def\T{{\cal{T}}}
\def\L{\Lambda}
\def\U{U_q(sl_2)}
\def\E{E_{\tau,\eta}(sl_n)}
\def\Zb{\mathbb{Z}}
\def\Cb{\mathbb{C}}

\def\R{\overline{R}}

\newcommand{\be}{\begin{eqnarray}}
\newcommand{\ee}{\end{eqnarray}}
\def\beq{\begin{equation}}
\def\eeq{\end{equation}}
\def\bea{\begin{eqnarray}}
\def\eea{\end{eqnarray}}
\def\ba{\begin{array}}
\def\ea{\end{array}}
\def\no{\nonumber}
\def\le{\langle}
\def\re{\}}
\def\lt{\left}
\def\rt{\right}
\def\non{\nonumber}

\newtheorem{Theorem}{Theorem}
\newtheorem{Definition}{Definition}
\newtheorem{Proposition}{Proposition}
\newtheorem{Lemma}{Lemma}
\newtheorem{Corollary}{Corollary}
\newcommand{\proof}[1]{{\bf Proof. }
        #1\begin{flushright}$\Box$\end{flushright}}

\title[]{On the complete-spectrum characterization of quantum integrable spin chains via inhomogeneous $T-Q$ relation}

\author{Junpeng Cao$^{1, 2}$, Wen-Li Yang$^{3, 4}$, Kangjie Shi$^{3}$ and Yupeng
Wang$^{1, 2}$}

\address{$^1$ Beijing National Laboratory for Condensed Matter Physics, Institute of Physics, Chinese Academy of Sciences, Beijing 100190, China\\
$^2$ Collaborative Innovation Center of Quantum Matter, Beijing,
     China\\
     $^3$ Institute of Modern Physics, Northwest University,
     Xian 710069, China\\
     $^4$ Beijing Center for Mathematics and Information Interdisciplinary Sciences, Beijing, 100048, China}

\ead{wlyang@nwu.edu.cn(Wen-Li Yang); yupeng@iphy.ac.cn(Yupeng Wang)}
\vspace{10pt}
\begin{indented}
\item[]July 2015
\end{indented}

\begin{abstract}
With the XXZ spin chains as examples, we prove two theorems: (1) the
functional relations derived from the off-diagonal Bethe Ansatz
scheme are the sufficient and necessary conditions to characterize
the complete spectrum of the corresponding transfer matrix; (2) each
eigenvalue of the transfer matrix can be parameterized by a minimal
inhomogeneous $T-Q$ relation. These statements hold for both with
and without inhomogeneity. The proof can be generalized to other
finite-dimensional quantum integrable models.
\end{abstract}

%
%
%
%
%

\section{Introduction}
\label{intro} \setcounter{equation}{0}
Recently, a method for solving the eigenvalue problem of quantum
integrable models with generic integrable boundary conditions, i.e.,
the off-diagonal Bethe Ansatz (ODBA) was proposed in
\cite{Cao1,Cao1-1,Cao1-2,Cao1-3} and several long-standing models
\cite{Cao1,Cao1-1,Cao1-2,Cao1-3,lcysw14,ZCYSW14,
Cao14,Li14,Hao14,Cao14-1} have since been solved (for details, see
\cite{wang}). The central idea of the method is to construct a
proper $T-Q$ relation \cite{bax} with an extra off-diagonal (or
inhomogeneous) term  based on the functional relations among
eigenvalues of transfer matrices. However, there are still some
concerns about the completeness and uniqueness of the solutions in
this scheme \cite{Nep13-1,jiang,nicc14-1}. In fact, completeness of
Bethe Ansatz solutions or Bethe Ansatz equations (BAEs) for
integrable models has been a longstanding problem for many years and
some attention has still been paid very recently \cite{Nepo14,
Nepo14-1}. In this paper, we take the XXZ spin chains (especially
the $U(1)$-symmetry-broken ones) as examples to prove that the
functional relations (necessary conditions that the eigenvalues of
the transfer matrix obey) derived in the framework of ODBA are also
sufficient conditions to determine the eigenvalues of the transfer
matrix. Therefore, those functional relations completely
characterize the spectrum of the transfer matrix in terms of certain
minimal inhomogeneous $T-Q$ relation, which gives rise to the Bethe
Ansatz solution to the underlying model.

The outline of the paper is  as follows. In the next section,  after
briefly reviewing the construction of the transfer matrices of the
XXZ spin-$\frac{1}{2}$ chain with generic open boundary conditions,
we give the direct proof that the solution set of the functional
relations obtained  via ODBA coincides exactly with the set of the
eigenvalues of the corresponding transfer matrix. In section 3, we
demonstrate  that each eigenvalue of the transfer matrix can be
parameterized by a polynomial $Q$-function in term of an
inhomogeneous $T-Q$ relation. Section 4 is attributed to the case of
open spin-$s$ chain. Concluding remarks and some discussion are
given in section 5.

\section{Complete spectrum of the transfer matrix}
\label{aniso} \setcounter{equation}{0}

The XXZ spin-$\frac{1}{2}$ chain with arbitrary boundary fields is
described by the Hamiltonian \bea
&&H=\sum_{j=1}^{N-1}\left[\sigma_{j}^{x}\sigma_{j+1}^{x}+\sigma_{j}^{y}
\sigma_{j+1}^{y}+\cosh\eta\sigma_{j}^{z} \sigma_{j+1}^{z}\right]
\nonumber \\
&&\; +\frac{ \sinh\eta}{\sinh\alpha_-\cosh\beta_-}
(\cosh\alpha_-\sinh\beta_- \sigma_{1}^z + \cosh\theta_-\sigma_1^x +
i\sinh\theta_- \sigma_1^y) \nonumber \\
&& \; + \frac{ \sinh\eta}{\sinh\alpha_+\cosh\beta_+}
(-\cosh\alpha_+\sinh\beta_+ \sigma_{N}^z + \cosh\theta_+\sigma_N^x +
i\sinh\theta_+ \sigma_N^y), \label{XXZ-Open}
\end{eqnarray} where 6 boundary parameters $\a_{\pm}$, $\b_{\pm}$ and $\theta_{\pm}$ are related to
the boundary fields. The well-known six-vertex $R$-matrix $R(u)\in
{\rm End}(V\otimes V)$ (with $V$ a two-dimensional vector space)
reads \bea R(u) = \frac{1}{\sinh\eta}\left(
\begin{array}{cccc}
    \sinh(u + \eta) &0            &0           &0            \\
    0                 &\sinh u     & \sinh\eta  &0            \\
    0                 &\sinh\eta   &  \sinh u    &0            \\
    0                 &0            &0           &\sinh(u + \eta)
\end{array} \right).
\label{R-matrix11} \eea Here and below we adopt the standard
notations: for any matrix $A\in {\rm End}({\rm V})$, $A_j$ is an
embedding operator in the tensor space ${ V}\otimes {
V}\otimes\cdots$, which acts as $A$ on the $j$-th space and as
identity on the other factor spaces; $R_{ij}(u)$ is an embedding
operator of $R$-matrix in the tensor space, which acts as identity
on the factor spaces except for the $i$-th and $j$-th ones. The
corresponding transfer matrix is given by \cite{sklyan} \bea t(u)=
tr_0\{K_0^+(u)\,T_0(u)K^-_0(u)\hat{T}_0(u)\},\label{Transfer-open}
\eea where the one-row monodronomy matrices are given by
\begin{eqnarray}
T_0(u)&=&R_{N0}(u-\theta_N)\ldots R_{10}(u-\theta_1),\no\\
\hat{T}_0(u)&=&R_{01}(u+\theta_1)\ldots
R_{0N}(u+\theta_N),\label{Mon-2}
\end{eqnarray}
$\{\theta_j|j=1,\cdots, N\}$ are the generic inhomogeneity
parameters, and the $K$-matrices are given by \cite{de1,GZ} \bea
K^-(u)&=&\lt(\begin{array}{ll}K^-_{11}(u)&K^-_{12}(u)\\
K^-_{21}(u)&K^-_{22}(u)\end{array}\rt),\no\\
K^-_{11}(u)&=&2\lt(\sinh(\a_-)\cosh(\b_{-})\cosh(u)
+\cosh(\a_-)\sinh(\b_-)\sinh(u)\rt),\no\\
K^-_{22}(u)&=&2\lt(\sinh(\a_-)\cosh(\b_{-})\cosh(u)
-\cosh(\a_-)\sinh(\b_-)\sinh(u)\rt),\no\\
K^-_{12}(u)&=&e^{\theta_-}\sinh(2u),\quad
K^-_{21}(u)=e^{-\theta_-}\sinh(2u),\label{K-matrix}\eea and \bea
K^+(u)=\lt.K^-(-u-\eta)\rt|_{(\a_-,\b_-,\theta_-)\rightarrow
(-\a_+,-\b_+,\theta_+)}.\label{K-6-2} \eea The commuting property
$[t(u),t(v)]=0$ ensures the integrability of the model.

It was shown in \cite{Cao1-3} that for generic
$\{\theta_j|j=1,\ldots,N\}$ the transfer matrix given by
(\ref{Transfer-open}) for arbitrary boundary parameters satisfies
the properties: \footnote{The relations (see
(\ref{Eigen-Identity-6-1}) below) satisfied by the eigenvalues of
the transfer matrix  for the XXZ spin-$\frac{1}{2}$ open chain with
one general non-diagonal and one diagonal or triangular boundary
K-matrices was  previously obtained by the separation of variables
method \cite{Fad14}. The first proof of the relation
(\ref{Operator-id-3}) for arbitrary boundary parameters on the
operator level (which does not depend on the basis) was given in the
reference \cite{Cao1-3}.} \bea
t(\theta_j)\,t(\theta_j-\eta)={a}(\theta_j){d}(\theta_j-\eta)\times
{\rm id},
\label{Operator-id-3}\\
t(-u-\eta)=t(u),\quad t(u+i\pi)=t(u),\label{Periodic-6}\\
t(0)=-2^3\sinh\a_-\cosh\b_-\sinh\a_+\cosh\b_+\cosh\eta\,\no\\
 \qquad \quad \times
\prod_{l=1}^N\frac{\sinh(\eta-\theta_l)\,\sinh(\eta+\theta_l)}{\sinh^2\eta}\times
{\rm id},\\
t(\frac{i\pi}{2})=-2^3\cosh\a_-\sinh\b_-\cosh\a_+\sinh\b_+\cosh\eta\,\no\\
 \qquad \quad \times
\prod_{l=1}^N\frac{\sinh(\frac{i\pi}{2}+\theta_l+\eta)\sinh(\frac{i\pi}{2}+\theta_l-\eta)}{\sinh^2\eta}
\times {\rm id},\\
\lim_{u\rightarrow \pm\infty}
t(u)=-\frac{\cosh(\theta_--\theta_+)e^{\pm[(2N+4)u+(N+2)\eta]}}
{2^{2N+1}\sinh^{2N}\eta}\times {\rm id} +\ldots,\label{Tran-Asy-6}
\eea where the functions ${a}(u)$ and ${d}(u)$ are given by
\cite{Cao1-3} \bea {a}(u)=-2^2 \frac{\sinh(2u+2\eta)}{\sinh(2u+
\eta)}
\sinh(u - \a_-)\no\\
 \qquad \quad \times \cosh(u - \b_-) \sinh(u - \a_+)
\cosh(u - \b_+){A}(u),\label{bar-a}\\
{d}(u)={a}(-u-\eta),\quad {A}(u)=
\prod_{l=1}^N\frac{\sinh(u-\theta_l+\eta)\sinh(u+\theta_l+\eta)}
{\sinh^2(\eta)}. \label{A-function} \eea The above operator
relations lead to that the corresponding eigenvalue, denoted by
$\Lambda(u)$, of the transfer matrix enjoys the properties: \bea
\Lambda(\theta_j)\Lambda(\theta_j-\eta)={a}(\theta_j){d}(\theta_j-\eta),
\quad j=1,\ldots,N, \label{Eigen-Identity-6-1}\\
 \L(-u-\eta)=\L(u),\quad \L(u+i\pi)=\L(u),\label{crosing-Eign-6}\\[2pt]
 \L(0)=-2^3\sinh\a_-\cosh\b_-\sinh\a_+\cosh\b_+\cosh\eta\no \\ \qquad\qquad
 \times
\prod_{l=1}^N\frac{\sinh(\eta-\theta_l)\,\sinh(\eta+\theta_l)}{\sinh^2\eta},\label{Eigen-6-1}\\[2pt]
 \L(\frac{i\pi}{2})=-2^3\cosh\a_-\sinh\b_-\cosh\a_+\sinh\b_+\cosh\eta\,\no\\[2pt]
 \quad\quad\quad\quad \times \prod_{l=1}^N\frac{\sinh(\frac{i\pi}{2}+\theta_l+\eta)\,\sinh(\frac{i\pi}{2}+\theta_l-\eta)}{\sinh^2\eta},\label{Eigen-6-2}\\[2pt]
 \lim_{u\rightarrow
\pm\infty}\L(u)=-\frac{\cosh(\theta_--\theta_+)e^{\pm[(2N+4)u+(N+2)\eta]}}
{2^{2N+1}\sinh^{2N}\eta}+\ldots,\label{Eigen-Asy-6}\\
 \L(u) \mbox{, as an entire function of $u$, } \no \\
 \qquad \;\; \mbox{is a trigonometric
polynomial of degree $2N+4$}.\label{Eigen-Anal-6} \eea The analogue
of the above relations in the homogeneous limit reads: \bea &&
\frac{\partial^l}{\partial u^l}
\lt.\Lambda(u)\Lambda(u-\eta)\rt|_{u=0} \no \\
&& = \frac{\partial^l}{\partial
u^l}\lt.\lt\{{a}(u)|_{\theta_j=0}\,{d}(u)|_{\theta_j=0}\rt\}\rt|_{u=0},
\quad
 l=0,1,\ldots,N-1, \label{Eigen-Identity-Hom-6-1}\\[2pt]
&& \L(-u-\eta)=\L(u),\quad \L(u+i\pi)=\L(u),\label{crosing-Eign-Hom-6}\\[2pt]
&& \L(0)=-2^3\sinh\a_-\cosh\b_-\sinh\a_+\cosh\b_+\cosh\eta,\label{Eigen-Hom-6-1}\\[2pt]
&&
\L(\frac{i\pi}{2})=(-1)^{N+1}2^3\cosh\a_-\sinh\b_-\cosh\a_+\sinh\b_+\cosh\eta\,
\frac{\cosh^{2N}\eta}{\sinh^{2N}\eta},\\[2pt]
&& \lim_{u\rightarrow
\pm\infty}\L(u)=-\frac{\cosh(\theta_--\theta_+)e^{\pm[(2N+4)u+(N+2)\eta]}}
{2^{2N+1}\sinh^{2N}\eta}+\ldots, \label{Eigen-Asy-Hom-6}\\[2pt]
&& \L(u) \mbox{, as an entire function of $u$, }\no \\ && \qquad \;
 \mbox{ is a trigonometric polynomial of degree
$2N+4$}.\label{Eigen-Anal-Hom-6} \eea

\begin{Proposition}  \label{Open}
The relations (\ref{Eigen-Identity-6-1})-(\ref{Eigen-Anal-6})
completely characterize the spectrum of the transfer matrix given by
(\ref{Transfer-open}) for the inhomogeneous XXZ spin-$\frac{1}{2}$
open chain with the most generic non-diagonal $K$-matrices specified
by  (\ref{K-matrix}) and (\ref{K-6-2}).
\end{Proposition}

\noindent{\it Proof\/}. An important fact is that for generic $\eta$
the generic boundary fields break all the non-abelian symmetries of the
Hamiltonian and induce non-degeneracy of the spectrum. Since all the eigenvalues of the transfer matrix must belong to the solutions of (\ref{Eigen-Identity-6-1})-(\ref{Eigen-Anal-6}), we conclude that the number of unequal solutions of (\ref{Eigen-Identity-6-1})-(\ref{Eigen-Anal-6}) must be larger or equal to $2^N$ (dimension of the Hilbert space)\footnote{Note that this conclusion does not hold if there is some degeneracy in the spectrum.}. On the other hand, the equations
(\ref{crosing-Eign-6}) and (\ref{Eigen-Asy-6})-(\ref{Eigen-Anal-6})
imply that the solutions form a $N+3$-dimensional linear space and
each of the solutions can be  expressed  uniquely in terms of  $N+3$
unknown coefficients $\{\bar{I}_{i}|i=0,\ldots,N+2\}$ as follows
\footnote{One may adopt another basis $\{(\sinh
u\sinh(u+\eta))^{i}(\cosh u\cosh(u+\eta))^{N+2-i}|i=0,\ldots,
N+2\}$.} \bea
\Lambda(u)=\sum_{i=0}^{N+2}\bar{I}_i\,\lt(e^{(2N+4-2i)u}+e^{-(2N+4-2i)(u+\eta)}\rt).
\eea Substituting the above expression into
(\ref{Eigen-Identity-6-1}) and (\ref{Eigen-6-1})-(\ref{Eigen-Asy-6})
gives rise to  $N+3$ equations with respect to the $N+3$ unknown
coefficients $\{\bar{I}_{i}|i=0,\ldots,N+2\}$. Among the resulting
equations, $N$ of them are quadratic in terms of
$\{\bar{I}_{i}|i=0,\ldots,N+2\}$ and the other 3 equations are
linear in terms of  $\{\bar{I}_{i}|i=0,\ldots,N+2\}$. According to
the B{\'e}zout Theorem \cite{Gri94}\footnote{The B{\'e}zout Theorem states that given two algebraic curves X and Y, the maximum number of their common points is the product of their degrees.} , we conclude that the maximum
number of the solutions of
(\ref{Eigen-Identity-6-1})-(\ref{Eigen-Anal-6}) is $2^N\times
1^3=2^N$. Based on the above arguments we
conclude that the number of solutions is exactly $2^N$ and the solution set of
(\ref{Eigen-Identity-6-1})-(\ref{Eigen-Anal-6}) is exactly  the eigenvalue set
 of the transfer matrix.
~~~~~~~~~~~~~~~~~~~~~~~~~~~~~~~~~~~~~~~~~~~~~~~~~~~~~~~~~~~~~~~~~~~~~~~~~~~~~~~~~~~~~~~~~~~~~~~~~~~~~~~~~~~~$\square$

\vspace{0.6truecm}

\noindent Similarly, we have
\begin{Corollary} \label{Hom-o}
For the homogenous XXZ spin-$\frac{1}{2}$  open chain with the most
generic non-diagonal $K$-matrices specified by  (\ref{K-matrix}) and
(\ref{K-6-2}), the spectrum of the transfer matrix is completely
determined by the equations
(\ref{Eigen-Identity-Hom-6-1})-(\ref{Eigen-Anal-Hom-6}).
\end{Corollary}

\noindent It was shown \cite{Fad14} that for each solution to
(\ref{Eigen-Identity-6-1})-(\ref{Eigen-Anal-6}), only with
non-trivial inhomogeneity, one could construct
 the corresponding eigenstate (i.e., the SoV-type eigensate whose homogeneous limit is still unclear)  of the transfer matrix. However, the corresponding Bethe-type  eigensate, no matter with or without inhomogeneity,  can be constructed \cite{Cao14-state,Bel14,Cao15} associated with each solution to (\ref{Eigen-Identity-6-1})-(\ref{Eigen-Anal-6}) (or (\ref{Eigen-Identity-Hom-6-1})-(\ref{Eigen-Anal-Hom-6})).\footnote{It should be emphasized that the inhomogeneous
$T-Q$ relation formalism (see (\ref{T-Q-relation-2}) below) plays a
key role in constructing the Bethe-type eigenstates
\cite{Cao14-state, Bel14, Cao15}.}

\section{Inhomogeneous $T-Q$ relation} \label{anisoopen} \setcounter{equation}{0}
In this section, we  show that each eigenvalue of the transfer
matrix (the solution of
(\ref{Eigen-Identity-6-1})-(\ref{Eigen-Anal-6}))  can be   expressed
in terms of some inhomogeneous $T-Q$ relation proposed in
\cite{Cao1-3}.
\begin{Proposition} \label{T-Q-Open}
Each solution of (\ref{Eigen-Identity-6-1})-(\ref{Eigen-Anal-6}) can
be parameterized in terms of the inhomogeneous $T-Q$ relation \bea
\Lambda(u)Q(u)={a}(u)Q(u - \eta) + {d}(u)Q(u + \eta) \no \\ \qquad
\qquad \quad + 2c\sinh2u\sinh(2u + 2\eta){A}(u) {A}( - u -
\eta),\label{T-Q-relation-2} \eea with $Q(u)$ being a trigonometric
polynomial as \bea
Q(u)=\prod_{j=1}^N\frac{\sinh(u-\l_j)\sinh(u+\l_j+\eta)}{\sinh^2\eta},
\label{Q-function-2} \eea and the constant $c$ being given by \bea
c=\cosh(\a_-+\b_-+\a_++\b_++(1+N)\eta)-\cosh(\theta_--\theta_+).\label{c-constant}
\eea The $N$ parameters $\{\l_j\}$ satisfy the associated BAEs \bea
{a}(\l_j)Q(\l_j-\eta)+{d}(\l_j)Q(\l_j+\eta)+2c\,\sin2\l_j\sinh(2\l_j+2\eta)
\no \\ \qquad \times {A}(\l_j){A}(-\l_j-\eta)=0,\quad
j=1,\ldots,N.\label{BAE-2} \eea
\end{Proposition}

\noindent{\it Proof\/}. Let us introduce a function $f(u)$ which is
equal to the difference between the LHS and the RHS of
(\ref{T-Q-relation-2}), namely, \bea f(u)= \Lambda(u)Q(u)-{a}(u)Q(u
- \eta) \no \\  \qquad  \quad - {d}(u)Q(u + \eta) -
2c\sinh2u\sinh(2u + 2\eta){A}(u) {A}( - u - \eta). \eea The
relations (\ref{crosing-Eign-6}) and (\ref{Eigen-Anal-6}) allow us
to derive
 that $f(u)$ satisfies the properties:
\bea &&f(u+i\pi)=f(u),\\
 &&f(u) \mbox{, as a function of $u$,}\no \\ &&\qquad \;\; \mbox{is a trigonometrical polynomial of degree $4N+4$}.
\eea This implies that together with the crossing symmetry, the
function $f(u)$ is fixed by its values at any  $2N+3$ different
points. For each solution of
(\ref{Eigen-Identity-6-1})-(\ref{Eigen-Anal-6}), one can always
choose a $Q(u)$ of form (\ref{Q-function-2}) such that \bea
&&f(0)=f(\frac{i\pi}{2})=f(\infty)=0,\label{Eq-0}\\
&&f(\theta_j)=\L(\theta_j)Q(\theta_j)-a(\theta_j)Q(\theta_j-\eta)=0,\quad
j=1,\ldots, N,\label{Eq-1}\\
&&f(\theta_j-\eta)=\L(\theta_j-\eta)Q(\theta_j-\eta)-d(\theta_j-\eta)Q(\theta_j)=0,
j=1,\ldots, N,\label{Eq-2} \eea which means $f(u)=0$ or
(\ref{T-Q-relation-2}) is fulfilled. In fact, the relation
(\ref{Eq-0}) is automatically satisfied, while the remaining  $2N$
equations (\ref{Eq-1})-(\ref{Eq-2}) can be rewritten in terms of the
following $N$ two-components equations \bea \lt(\begin{array}{cc}
\L(\theta_j)&-a(\theta_j)\\
d(\theta_j-\eta)&\L(\theta_j-\eta)\end{array}\rt)
\lt(\begin{array}{c} Q(\theta_j)\\ Q(\theta_j-\eta)\end{array}\rt)=
\lt(\begin{array}{c} 0\\ 0\end{array}\rt),  j=1,\ldots,N. \eea The
conditions that the above $N$ equations have non-trivial solutions
is exact (\ref{Eigen-Identity-6-1}). This means that if $\L(u)$ is a
solution of (\ref{Eigen-Identity-6-1})-(\ref{Eigen-Anal-6})), the
$2N$ equations (\ref{Eq-1}) and (\ref{Eq-2}) are equivalent to $N$
equations \bea
 \L(\theta_j)Q(\theta_j)=a(\theta_j)Q(\theta_j-\eta),\quad j=1,\ldots,N,\label{Eq-3}
\eea which allow us to determine the $Q(u)$ function in the form of
(\ref{Q-function-2}) by its values at the $N$ points $\theta_j$.
Therefore, we are always able to choose the $Q(u)$ of form
(\ref{Q-function-2}) from  (\ref{Eq-3}) such that $f(u)=0$ provided
that $\Lambda(u)$ is an eigenvalue of the transfer matrix
(\ref{Transfer-open}). Moreover, taking $u$ at the roots of the
$Q(u)$ function (i.e., $\{\l_j\}$), the condition $f(\l_j)=0$ gives
rise to the
 associated BAEs (\ref{BAE-2}), which determine the $Q(u)$ function\footnote{Note that we only claim that each solution of $\L(u)$ can be parameterized by an inhomogeneous $T-Q$ relation. This does not mean that all $T-Q$ relations given by the solutions of the BAEs correspond to right eigenvalues because there may be ``unphysical" solutions to the BAEs as discussed in \cite{Nepo14,Nepo14-1}. Nevertheless we would point out that the ``irregular" solutions of the BAEs with $\lambda_j=\theta_j, \theta_j-\eta$ do not satisfy (\ref{Eigen-Identity-6-1}) in our case.}. This completes the proof.
~~~~~~~~~~~~~~~~~~~~~~~~~~~~~~~~~~~~~~~~~~~~~~~~~~~~~~~~~~~~~~~~~~~~~~~~~~~~~~~~~~~~~~~~~~~~~~~~~~~~~~~~~~~~$\square$

\vspace{0.6truecm}

\noindent Some remarks are in order. (1) Provided that all the
eigenvalues $\L(u)$ are simple, $N$ is the minimal degree of the
trigonometric polynomial $Q(u)$ (\ref{Q-function-2}) enabling one to
parameterize any eigenvalue $\Lambda(u)$ in the form of
(\ref{T-Q-relation-2}). (2) Actually there exist an infinite number
of possible ways \cite{Cao1-3} to parameterize  the eigenvalue of
the transfer matrix, but they are all equivalent to each other
because of the finite number of eigenvalues. The corresponding Bethe
states for different parametrization were constructed
\cite{Cao14-state, Bel14, Cao15}. (3) The degree of the
$Q$-polynomial may be reduced to a small value for the case that the
inhomogeneous term vanishes. In this case the $T-Q$ relation becomes
a homogeneous one (the well-known Baxter's $T-Q$ relation). This
happens in cases of the $U(1)$ symmetry or in the degenerate cases
of the open spin chain \cite{CLSW}, where the transfer matrix can be
diagonalized in some smaller blocks.

\section{Results for the XXZ spin-$s$ open chain}\setcounter{equation}{0}

For the XXZ spin-$s$ chain
\cite{Zam80,fusion,fusion-1,fusion-2,fusion-3,MNR}
($s=\frac{1}{2},\,1,\,\frac{3}{2},\ldots$), the quantum space of
each site is $2s+1$-dimensional and endows the spin-$s$
representation of $U_q(sl_2)$ with $q=e^{\eta}$ \cite{Cha94}.  The
fundamental spin-$(\frac{1}{2},s)$ $R$-matrix is given by
\cite{fusion,fusion-1,fusion-2,fusion-3} \bea
R^{(\frac{1}{2},s)}_{12}(u)&=&\frac{1}{\sinh\eta}\lt(\begin{array}{cc}\sinh(u+\frac{\eta}{2}+\eta
S^3_2)&\sinh\eta \,\,S^-_2\\ \sinh\eta\,\,
S^+_2&\sinh(u+\frac{\eta}{2}-\eta S^3_2)\end{array}\rt),\label{R-1s}
\eea  where ${S^3,\,S^{\pm}}$ are the spin-$s$ realizations of the
quantum algebra $U_q(sl_2)$. The fundamental transfer matrix
 denoted by $t^{(\frac{1}{2},s)}(u)$,  is given by (cf. (\ref{Transfer-open}))
\bea
t^{(\frac{1}{2},s)}(u)=tr_0\{K_0^+(u)\,T^{(\frac{1}{2},s)}_0(u)K^-_0(u)\hat{T}^{(\frac{1}{2},s)}_0(u)\},\no
\eea where the two monodromy matrices are given by \bea
 T^{(\frac{1}{2},s)}(u)&=&R^{(\frac{1}{2},s)}_{0N}(u-\theta_N)\ldots
 R^{(\frac{1}{2},s)}_{01}(u-\theta_1),\no\\
 \hat{T}^{(\frac{1}{2},s)}(u)&=&R^{(\frac{1}{2},s)}_{01}(u+\theta_1)\ldots R^{(\frac{1}{2},s)}_{0N}(u+\theta_N),\no
\eea and the corresponding $K$-matrices are given by
(\ref{K-matrix}) and (\ref{K-6-2}) respectively. The transfer matrix
and its fused ones (for details we refer to
\cite{Nep04,Nep04-1,Nep04-2,Nep04-3}), denoted by $t^{(j,s)}(u)$,
form the commutative families, which ensures the integrability of
the corresponding model. Moreover these satisfy the fusion hierarchy \cite{kulish}
relations \cite{fusion2,fusion2-1,Yan06,Fra07} \bea
t^{(\frac{1}{2},s)}(u)\, t^{(j-\frac{1}{2},s)}(u- j\eta)=
t^{(j,s)}(u-(j-\frac{1}{2})\eta) \no \\ \qquad \quad  +
\delta^{(s)}(u)\, t^{(j-1,s)}(u-(j+\frac{1}{2})\eta),\quad j =\frac
12, 1, \frac{3}{2}, \ldots, \label{Hier-Op-s} \eea where we have
used the convention $t^{(0,s)}={\rm id}$ and the coefficient
function $\delta^{(s)}(u)$ related to the quantum determinant is
given by \bea &&
\d^{(s)}(u)=\bar{a}^{(s)}(u)d^{(s)}(u-\eta),\label{d-s}\\ &&
\bar{a}^{(s)}(u)= - 2^2 \frac{\sinh(2u + 2\eta)}{\sinh(2u + \eta)}
\sinh(u - \a_-)\cosh(u - \b_-) \no \\ && \qquad\qquad \times \sinh(u
- \a_+) \cosh(u - \b_+)\bar{A}^{(s)}(u),\label{bar-a-s}\\
&& \bar{d}^{(s)}(u)=\bar{a}^{(s)}( - u - \eta), \\
&&\bar{A}^{(s)}(u)=
\prod_{l=1}^N\frac{\sinh(u-\theta_l+(\frac{1}{2}+s)\eta)\sinh(u+\theta_l+(\frac{1}{2}+s)\eta)}
{\sinh^2(\eta)}. \label{A-S-function} \eea The fused transfer
matrices $t^{(j,s)}(u)$, which are mutually commuting, can be
constructed by the fused $R$-matrices and $K$-matrices
\cite{fusion,MNR,fusion2,fusion2-1,Fra07}. Let us denote the
eigenvalue of the transfer matrix $t^{(j,s)}(u)$ by
$\Lambda^{(j,s)}(u)$. The commutativity of the transfer matrices and
the operator relations (\ref{Hier-Op-s}) imply that the eigenvalues
also obey the same fusion hierarchy relations \bea
\Lambda^{(\frac{1}{2},s)}(u)\, \Lambda^{(j-\frac{1}{2},s)}(u-
j\eta)= \Lambda^{(j,s)}(u-(j-\frac{1}{2})\eta) \no \\ \qquad \quad +
\delta^{(s)}(u)\, \Lambda^{(j-1,s)}(u-(j+\frac{1}{2})\eta),\quad j
=\frac 12, 1, \frac{3}{2}, \ldots. \label{Hier-Eigen-s} \eea
Following the method developed in \cite{Cao14, Hao14}, we can obtain
some operator identities between $t^{(\frac{1}{2},s)}(u)$ and
$t^{(s,s)}(u)$ at some special points, which leads to the following
relation among their eigenvalues \cite{Cao14-1,ap} \bea
\Lambda^{(s,s)}(\theta_j)
\,\Lambda^{(\frac{1}{2},s)}(\theta_j-(\frac{1}{2}+s)\eta)\no
\\ =\d^{(s)}(\theta_j+(\frac{1}{2}-s)\eta)\,
\Lambda^{(s-\frac{1}{2},s)}(\theta_j+\frac{\eta}{2}),\quad
j=1,\ldots,N.\label{Eigen-Identity-s} \eea Moreover, it is  easy to
verify that $\Lambda^{(\frac{1}{2},s)}(u)$ satisfy the relations:
\bea
&& \L^{(\frac{1}{2},s)}(u+i\pi)=\L^{(\frac{1}{2},s)}(u),\quad \L^{(\frac{1}{2},s)}(-u-\eta)=\L^{(\frac{1}{2},s)}(u),\label{Eigenvalue-crossing-s}\\
&& \L^{(\frac{1}{2},s)}(0)=-2^3\sinh\a_-\cosh\b_-\sinh\a_+\cosh\b_+\cosh\eta\,\no\\
&& \quad\quad\quad\quad\quad\quad\times
\prod_{l=1}^N\sinh(\theta_{l}+(\frac{1}{2}+s)\eta)\sinh(-\theta_{l}+(\frac{1}{2}+s)\eta),
 \label{Eigenvalue-Int-1-s}\\
&& \L^{(\frac{1}{2},s)}(\frac{i\pi}{2})=-2^3\cosh\a_-\sinh\b_-\cosh\a_+\sinh\b_+\cosh\eta\,\no\\
&&
\quad\quad\quad\quad\times\prod_{l=1}^N\sinh(\frac{i\pi}{2}+\theta_{l}+(\frac{1}{2}+s)\eta)
\sinh(\frac{i\pi}{2}+\theta_{l}-(\frac{1}{2}+s)\eta),
 \label{Eigenvalue-Int-2-s}\\
&& \L^{(\frac{1}{2},s)}(u)\lt|_{u\rightarrow\pm\infty}\rt.=
-\frac{\cosh(\theta_--\theta_+)e^{\pm[(2N+4)u+(N+2)\eta]}}
{2^{2N+1}\sinh^{2N}\eta} +\ldots,\label{Eigenvalue-Asy-s}\\
&& \L^{(\frac{1}{2},s)}(u) \mbox{, as a function of $u$,}\no \\
&& \qquad \quad \;\; \mbox{ is a trigonometric polynomial of degree
$2N+4$}.\label{Eigenvalue-Anal-s} \eea The equations
(\ref{Eigenvalue-crossing-s}),
(\ref{Eigenvalue-Asy-s})-(\ref{Eigenvalue-Anal-s}) allow us to
express  $\L^{(\frac{1}{2},s)}(u)$ uniquely in terms of  $N+3$
unknown coefficients $\{\bar{I}^{(s)}_{i}|i=0,\ldots,N+2\}$ as
follow \bea
\Lambda^{(\frac{1}{2},s)}(u)=\sum_{i=0}^{N+2}\bar{I}^{(s)}_i\,\lt(e^{(2N+4-2i)u}+e^{-(2N+4-2i)(u+\eta)}\rt).\no
\eea The hierarchy relation  (\ref{Hier-Eigen-s}) implies that
$\Lambda^{(s,s)}(u)$, as a function of
$\{\bar{I}^{(s)}_{i}|i=0,\ldots,N+2\}$, is a polynomial of degree
$2s$. Then each equation of (\ref{Eigen-Identity-s}), with respect
to $\{\bar{I}^{(s)}_{i}|i=0,\ldots,N+2\}$, is of degree $2s+1$.
Noting the fact that the dimension of the Hilbert space of the
spin-$s$ chain is $(2s+1)^N$, with similar analysis for the
spin-$\frac12$ case we conclude that the equations
(\ref{Hier-Eigen-s})-(\ref{Eigenvalue-Anal-s}) completely
characterize the spectrum of the transfer matrix
$t^{(\frac{1}{2},s)}(u)$, and that each eigenvalue
$\Lambda^{(\frac{1}{2},s)}(u)$ of the transfer matrix
$t^{(\frac{1}{2},s)}(u)$ can be parameterized by the inhomogeneous
$T-Q$ relation \bea
 \L^{(\frac{1}{2},s)}(u)&=&a^{(s)}(u)\frac{Q(u-\eta)}{Q(u)}
                           +d^{(s)}(u)\frac{Q(u+\eta)}{Q(u)}\no\\
 &&+2c^{(s)}\,\sinh 2u\,\sinh(2u+2\eta)\frac{F^{(s)}(u)}{Q(u)},\label{T-Q-Ansatz-s}
\eea where  the function $F^{(s)}(u)$ and the constant $c^{(s)}$ are
given by \bea
F^{(s)}(u)=\prod_{j=1}^N\prod_{k=0}^{2s}\frac{\sinh(u-\theta_{j}+(\frac{1}{2}-s+k)\eta)
\sinh(u+\theta_{j}+(\frac{1}{2}-s+k)\eta)}{\sinh\eta\,\sinh\eta},\no\\
c^{(s)}=\cosh(\a_-+\b_-+\a_++\b_++(1+2sN)\eta)-\cosh(\theta_--\theta_+).\label{c-constant-s}
\eea The  function $Q(u)$ is parameterized by $2sN$ Bethe roots
$\{\l_j|j=1,\ldots,2sN\}$ as \bea
Q(u)&=&\prod_{j=1}^{2sN}\frac{\sinh(u-\l_j)\,\sinh(u+\l_j+\eta)}{\sinh\eta\,\sinh\eta}=Q(-u-\eta),\no
\eea which should satisfy the BAEs \bea a^{(s)}(\l_j)Q(\l_j-\eta)+
d^{(s)}(\l_j)Q(\l_j+\eta) \no \\[2pt] \qquad +2c^{(s)}\sinh
2\l_j\sinh(2\l_j+2\eta)F^{(s)}(\l_j)=0,\quad
 j=1,\ldots,2sN.\label{BAEs-s}
\eea

\section{Conclusions}\setcounter{equation}{0}

The complete-spectrum characterization of quantum integrable spin
chains via ODBA is studied. Taking the XXZ spin chains (especially
for the $U(1)$-symmetry-broken ones) as examples, we have shown that
the relations (\ref{Eigen-Identity-6-1})-(\ref{Eigen-Anal-6})
completely characterize the spectrum of the transfer matrix of the
spin-$\frac{1}{2}$ XXZ chain with generic non-diagonal open boundary
condition. It is further shown that each eigenvalue of the transfer
matrices can  be parameterized in terms of some inhomogeneous $T-Q$
relation such as (\ref{T-Q-relation-2}), which gives rise to the
Bethe Ansatz solution to the underlying model \cite{Cao1}. Moreover,
for the spin-$s$ chain, the relations
(\ref{Hier-Eigen-s})-(\ref{Eigenvalue-Anal-s}) completely
characterize the spectrum of the fundamental  transfer matrix
$t^{(\frac{1}{2},s)}(u)$ and each eigenvalue can be given in terms
of the inhomogeneous $T-Q$ relation (\ref{T-Q-Ansatz-s}).

The inhomogeneous $T-Q$ relation of type (\ref{T-Q-relation-2}) or
(\ref{T-Q-Ansatz-s}), like the role of the Yang-Baxter equation,
plays  a unified and fundamental role to characterize the eigenvalue
of the transfer matrix of rank-one quantum integrable models both
with and without $U(1)$ symmetry.

\section*{Acknowledgments}

We would like to thank F. Gong for drawing our attention to the
B{\'e}zout Theorem. The financial supports from the National Natural
Science Foundation of China (Grant Nos. 11375141, 11374334, 11434013
and 11425522), the National Program for Basic Research of MOST (973
project under grant No. 2011CB921700) and BCMIIS are gratefully
acknowledged. Two of the authors (W.-L. Yang and K. Shi) would like
to thank IoP, CAS for the hospitality. W.-L. Yang also would like to
thank KITPC for the hospitality where some part of the work was done
during his visiting.

\end{document}